\documentclass[10pt,a4paper]{article}
\usepackage[T1]{fontenc}
\usepackage{microtype}
\usepackage{amsmath,amssymb}
\usepackage{amsfonts}
\usepackage{mathrsfs}
\usepackage[table]{xcolor}  
\usepackage{booktabs}
\usepackage{array}
\usepackage{multirow}
\usepackage{longtable}
\usepackage{tabularx}
\usepackage{xltabular}

\usepackage{graphicx}
\usepackage[export]{adjustbox}
\usepackage{pdflscape}
\usepackage{rotating}
\usepackage{caption}
\usepackage{subcaption}
\usepackage{authblk}
\usepackage{etoolbox}
\usepackage{orcidlink}
\usepackage{textcomp}
\usepackage[margin=1.5cm,top=1.5cm]{geometry}
\usepackage{hyperref}
\usepackage[numbers,sort&compress]{natbib}
\hypersetup{colorlinks=true, linkcolor=blue, citecolor=blue, urlcolor=blue}
\providecommand{\keywords}[1]{%
  \par\vspace{0.5ex}%
  \noindent\textbf{Keywords: }#1\par
}

\numberwithin{equation}{section}
\numberwithin{figure}{section}
\numberwithin{table}{section}
\newcolumntype{C}{>{\centering\arraybackslash}X}
\newcolumntype{L}{>{\raggedright\arraybackslash}X}
\newcolumntype{R}{>{\raggedleft\arraybackslash}X}
\renewcommand{\arraystretch}{1.3}
\begin{document}

\begin{center}
\LARGE{Periodic Orbits and Gravitational Wave Signatures around the Bonanno--Reuter Regular Black Hole \\\;}
\par\end{center}

\begin{center}
{\bf Mohammad Reza Alipour\orcidlink{0000-0001-8074-7865}}\footnote{\bf mohamad.alipour.1994@gmail.com; mr.alipour@stu.umz.ac.ir}\\
{\it School of Physics, Damghan University, P.~O.~Box 3671641167, Damghan, Iran}\\
{\it Center for Theoretical Physics, Khazar University, 41 Mehseti Street, Baku, AZ1096, Azerbaijan}
\end{center}

\begin{center}
{\bf Mohammad Ali S. Afshar\orcidlink{0009-0001-3133-5992}}\footnote{\bf m.a.s.afshar@gmail.com}\\
{\it Department of Physics, Faculty of Basic Sciences, University of Mazandaran P. O. Box 47416-95447, Babolsar, Iran}\\
{\it School of Physics, Damghan University, P.~O.~Box 3671641167, Damghan, Iran}\\
{\it Center for Theoretical Physics, Khazar University, 41 Mehseti Street, Baku, AZ1096, Azerbaijan}
\end{center}

\begin{center}
{\bf Saeed Noori Gashti\orcidlink{0000-0001-7844-2640}}\footnote{\bf sn.gashti@du.ac.ir; saeed.noorigashti70@gmail.com}\\
{\it School of Physics, Damghan University, P.~O.~Box 3671641167, Damghan, Iran}
\end{center}

\begin{center}
{\bf Behnam Pourhassan\orcidlink{0000-0003-1338-7083}}\footnote{\bf b.pourhassan@du.ac.ir}\\
{\it School of Physics, Damghan University, P.~O.~Box 3671641167, Damghan, Iran}\\
{\it Center for Theoretical Physics, Khazar University, 41 Mehseti Street, Baku, AZ1096, Azerbaijan}
\end{center}

\begin{center}
{\bf Jafar Sadeghi \orcidlink{0000-0003-4549-1766}}\footnote{\bf pouriya@ipm.ir}\\
{\it Department of Physics, Faculty of Basic Sciences, University of Mazandaran P. O. Box 47416-95447, Babolsar, Iran}\\
{\it School of Physics, Damghan University, P.~O.~Box 3671641167, Damghan, Iran}
\end{center}

\begin{abstract}
Timelike geodesics, periodic orbits, and their associated gravitational-wave signatures are examined in the spacetime of a Bonanno--Reuter regular black hole, a geometry arising from Asymptotically Safe Gravity in which a running Newton coupling replaces the central singularity with a de Sitter core. The dimensionless parameter $\alpha/M^2$ completely determines the strong-field dynamics. Increasing $\alpha/M^2$ shifts the marginally bound and innermost stable circular orbits inward, systematically reducing their characteristic radii, angular momenta, and energies; the allowed phase space for bound motion contracts accordingly. Classifying trajectories via the rational frequency ratio $q = w + v/z$ reveals that periodic orbits experience a mild inward contraction, which reduces the energy necessary to sustain a specific topology. Within the numerical kludge framework, we calculate the gravitational-wave polarizations for extreme mass-ratio inspirals. The asymptotically safe correction induces a leftward phase shift that reflects shorter orbital periods, while mildly enhancing peak amplitudes owing to the smaller periastron distances reached in the deep strong-field regime. Waveform sensitivity displays a strong dependence on topology, with high-whirl orbits, which persist longer in the strong-field region near the horizon, showing markedly more pronounced deviations. Unlike environmental effects that inflate orbital scales, intrinsic quantum-gravity modifications generate distinct, observationally detectable signatures for future space-based detectors such as LISA, Taiji, and TianQin.
\end{abstract}


\tableofcontents

\section{Introduction}
\label{sec:intro}

General relativity (GR)~\cite{Einstein1916} provides a robust description of black holes (BHs)~\cite{Schwarzschild1916,Kerr1963}, which serve as premier laboratories for strong-field gravity. The direct detection of gravitational waves (GWs) from compact-binary coalescences~\cite{AbbottGW150914,AbbottGW151226} and the horizon-scale imaging of supermassive objects in M87$^{\ast}$ and Sgr~A$^{\ast}$~\cite{EHTM87,EHTSgrA} have validated GR in its most extreme regimes. Yet rigorous strong-field tests remain essential for understanding horizon structure, GW physics, and the nature of gravity itself~\cite{Will2014,Psaltis2020}.

Future space-based detectors such as LISA, Taiji, and TianQin~\cite{AmaroSeoaneLISA,HuWuTaiji2017,GongTianQin2021} will be sensitive to low-frequency sources, most notably extreme mass-ratio inspirals (EMRIs). In an EMRI, a stellar-mass compact object orbits a supermassive BH, emitting long-lived GWs whose waveforms encode detailed information about the underlying spacetime geometry. Because any change in the background spacetime modifies orbital frequencies, zoom-whirl dynamics, and the resulting EMRI signal, EMRIs serve as sensitive probes of strong-field dynamics~\cite{Hughes2001,AmaroSeoaneEMRI,Babak2017,GlampedakisKennefick2002,RuangsriHughes2014,BarackEtAl2019}.

The gravitational waveforms emitted by EMRIs are fundamentally shaped by \emph{periodic orbits}~\cite{GrossmanLevin2009,MisraLevin2010}. These bound trajectories close upon themselves after a finite number of radial ($\omega_{r}$) and azimuthal ($\omega_{\phi}$) oscillations. Following the taxonomy introduced by Levin and Perez-Giz, each periodic orbit is uniquely characterized by a rational frequency ratio $\omega_{\phi}/\omega_{r}$ and labeled by three integers $(z,w,v)$---the zoom, whirl, and vertex numbers~\cite{LevinPerezGiz2008,Levin2009}. Originally developed for Schwarzschild and Kerr spacetimes~\cite{BambhaniyaEtAl2021,RanaMangalam2019}---and together with the zoom-whirl behavior first mapped for black-hole binaries~\cite{HealyLevinShoemaker2009}—this framework has since been applied to a wide variety of backgrounds, including the $\gamma$-metric~\cite{ZhangZhuGamma2026}, quantum-corrected and regular geometries~\cite{ChenYang2025,LiKuang2026,HuaEtAl2026,KumarMohammadiGhosh2026}, alternative gravity theories~\cite{LuZhu2025,JuniorEtAl2025,WangCauchy2025}, non-commutative black holes~\cite{HeidariEtAl2026,AhmedQuintessence2026,UktamovEtAl2026}, and BHs embedded in dark matter halos~\cite{ShokirovEtAl2026,HeidariAraujoFilho2026Hernquist,AlloqulovDehnen2025,HaroonZhu2025}. These studies predominantly employ the numerical kludge scheme~\cite{BabakKludge2007,PoissonWill2014}, combining geodesic integration with the quadrupole formula.

A systematic dichotomy emerges from this body of work: environmental effects tend to enlarge bound motion regions and inflate periodic orbits, whereas intrinsic corrections---such as charge or short-distance quantum effects---act in the opposite direction, contracting orbits and lowering their characteristic energies and angular momenta. It is within this latter context of intrinsic modifications that the present study is naturally situated.

A closely related recent study is that of Kumar et al.~\cite{KumarMohammadiGhosh2026}, 
who analyzed periodic orbits and EMRI waveforms for a different ASG-motivated regular black hole, 
derived from a dust-collapse model with a logarithmic lapse function whose quantum correction falls 
off as $r^{-4}$. The Bonanno--Reuter geometry considered here is constructed differently---directly 
from an RG-improvement of the Schwarzschild solution---and leads to a rational lapse function whose 
leading correction decays more slowly, as $r^{-3}$ (originating from a running coupling $G(r)$ that 
falls off as $r^{-2}$), while replacing the singularity with a genuine finite-curvature de Sitter 
core~\cite{BaticDutykhScardigli2026}. The two models therefore differ in origin, functional form, 
asymptotic behavior, and the allowed range of their respective quantum parameters ($\xi$ vs.~$\alpha/M^{2}$), 
so their imprints on the orbital and waveform structure are not expected to coincide, even though 
both employ the same Levin--Perez-Giz taxonomy and numerical-kludge framework. The present work 
extends this line of study to the original, historically foundational Bonanno--Reuter black 
hole~\cite{BonannoReuter2000}, whose periodic-orbit and EMRI phenomenology has not previously been 
derived.

In this work, we fill this gap. We demonstrate that the strong-field dynamics is governed entirely by the dimensionless parameter $\alpha/M^{2}$. We determine how the marginally bound orbit (MBO) and the innermost stable circular orbit (ISCO) migrate as the ASG correction is switched on, mapping the bound-orbit window in the $(E,L)$ plane. We then construct periodic orbits across the $(z,w,v)$ taxonomy, quantifying the systematic inward contraction induced by ASG. Finally, within the numerical kludge framework, we compute the time-domain $h_{+}$ and $h_{\times}$ polarizations for a supermassive primary and assess their detectability with LISA, Taiji, and TianQin.

The remainder of the paper is organized as follows. Section~\ref{sec:metric} introduces the Bonanno--Reuter metric. Section~\ref{sec:mbo-isco} develops the timelike geodesic equations and characteristic orbits. Section~\ref{sec:periodic} classifies periodic orbits and analyzes their zoom-whirl morphology. Section~\ref{sec:waveforms} presents the GW waveforms and discusses their detectability. We work throughout in geometrized Planck units, $c=k_{B}=G_{N}=\hbar=1$, and adopt the mostly-plus metric signature.

\section{Spacetime Metric and Geodesic Structure}
\label{sec:metric}

The Bonanno--Reuter black hole is a regular geometry motivated by Asymptotically Safe Gravity (ASG)~\cite{BonannoReuter2000}, wherein Newton's constant becomes a running coupling $G(k)$ that vanishes at the non-Gaussian ultraviolet fixed point. Promoting this to a position-dependent coupling $G(r)$ via a cutoff identification $k(r)$ and substituting it into the Schwarzschild solution yields a renormalization-group-improved metric. In geometrized Planck units ($c=k_B=G_N=\hbar=1$),
\begin{equation}
G(r) = \frac{r^{3}}{r^{3}+\alpha\,(r+\gamma M)},
\label{eq:Grunning}
\end{equation}
where $\alpha>0$ ensures asymptotic safety, and $\gamma$ originates from the cutoff. We adopt the standard values $\alpha=118/(15\pi)$ and $\gamma=9/2$ throughout, following Refs.~\cite{BonannoReuter2000,BaticDutykhScardigli2026}.

This yields the static, spherically symmetric line element
\begin{equation}
ds^{2} = -f(r)\,dt^{2} + \frac{dr^{2}}{f(r)} + r^{2}\left(d\theta^{2}+\sin^{2}\theta\,d\phi^{2}\right),
\label{eq:metric}
\end{equation}
with the RG-improved lapse function
\begin{equation}
f(r) = 1-\frac{2M\,r^{2}}{D(r)}, \qquad D(r) \equiv r^{3}+\alpha\,r+\alpha\gamma M.
\label{eq:lapse}
\end{equation}
The metric interpolates between the classical Schwarzschild solution at large radii ($D(r) \to r^3$) and a regular de Sitter core as $r \to 0$, where $f(r) \simeq 1 - 2r^2/(\alpha\gamma)$. Consequently, the central curvature singularity is replaced by a finite-curvature center, and the Kretschmann scalar remains finite everywhere for $\alpha,\gamma>0$~\cite{BaticDutykhScardigli2026}.

The horizon structure is modified accordingly. The horizons are determined by the positive roots of $D(r)-2Mr^2=0$, specifically
\begin{equation}
r_{h}^{3} - 2M\,r_{h}^{2} + \alpha\,r_{h} + M\gamma\alpha = 0.
\label{eq:horizon-eq}
\end{equation}
For $M > M_c$, this cubic admits two positive roots $r_- < r_+$, describing a non-extremal black hole with outer event horizon $r_h \equiv r_+$. At the critical mass $M = M_c$, these roots degenerate into a single extremal horizon $r_e$, yielding numerically $M_c \simeq 3.502741812$ and $r_e \simeq 4.484183919$~\cite{BaticDutykhScardigli2026}. Unless stated otherwise, we operate on the non-extremal branch $M > M_c$.

Because $\alpha$ and $M$ share the same dimensions, the entire strong-field dynamics is governed by the single dimensionless ratio $\alpha/M^{2}$. All characteristic orbital quantities depend solely on this parameter, bounded above by the critical value $\alpha/M_c^{2} \simeq 0.204091361$. Moreover, since $D(r) > 0$ for all $r > 0$, the lapse function remains smooth and pole-free outside the event horizon, ensuring the numerical stability of the geodesic analysis in the Sec.~\ref{sec:mbo-isco}.

\section{Timelike Geodesics and Characteristic Orbits}
\label{sec:mbo-isco}

\begin{figure}[htbp]
\centering
\includegraphics[width=0.72\linewidth]{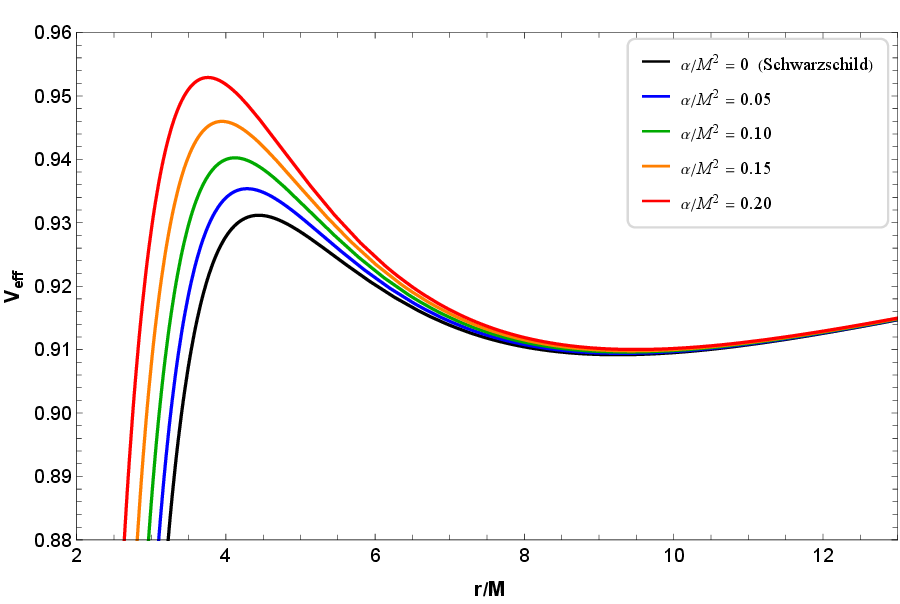}
\caption{Effective potential $V_{\rm eff}$ for fixed angular momentum $L/M=3.7$ and several values of $\alpha/M^{2}$, with $\gamma=9/2$. The black curve ($\alpha/M^2=0$) is the Schwarzschild case.}
\label{fig:veff-profiles}
\end{figure}

For a massive test particle confined to the equatorial plane ($\theta=\pi/2$), the geodesic Lagrangian
\begin{equation}
\mathcal{L} = \frac{1}{2}\,g_{\mu\nu}\dot{x}^{\mu}\dot{x}^{\nu}
            = \frac{1}{2}\left(-f(r)\,\dot t^{2}
              + \frac{\dot r^{2}}{f(r)} + r^{2}\dot\phi^{2}\right)
\label{eq:lagrangian}
\end{equation}
yields the conserved specific energy $E = f(r)\,\dot t$ and angular momentum $L = r^{2}\,\dot\phi$. Using the timelike normalization $g_{\mu\nu}\dot x^\mu\dot x^\nu=-1$, the radial dynamics reduce to $\dot r^{2} = E^{2} - V_{\rm eff}(r)$, with the effective potential
\begin{equation}
V_{\rm eff}(r) = f(r)\left(1+\frac{L^{2}}{r^{2}}\right) = \left(1-\frac{2M r^{2}}{D(r)}\right)\left(1+\frac{L^{2}}{r^{2}}\right).
\label{eq:veff}
\end{equation}
Since $D(r)>0$ outside the horizon, $V_{\rm eff}$ is smooth for all $r>r_h$. Furthermore, because $\alpha$ and $M$ enter only through the dimensionless ratio $\alpha/M^{2}$, this parameter solely governs the strong-field dynamics. As shown in Fig.~\ref{fig:veff-profiles} for fixed $L/M=3.7$, increasing $\alpha/M^{2}$ raises the potential barrier and shifts it inward. Asymptotically, the $1/r^{3}$ decay of the ASG correction ensures all curves converge to the Schwarzschild profile at large radii.

The domain of bound motion is bounded by the marginally bound orbit (MBO) and the innermost stable circular orbit (ISCO). For static, spherically symmetric metrics satisfying $g_{tt}g_{rr}=-1$, the circular orbit condition $\partial_r V_{\rm eff}=0$ yields $L^{2}(r) = r^{3}f'(r)/[2f(r)-r f'(r)]$. Substituting this back into $V_{\rm eff}$ reduces the MBO and ISCO conditions to single equations in $r$. The MBO, defined by $V_{\rm eff}(r_{\rm MBO})=1$ and $\partial_r V_{\rm eff}=0$, gives
\begin{equation}
r_{\rm MBO} = \frac{2\,f(r_{\rm MBO})\left[1-f(r_{\rm MBO})\right]}{f'(r_{\rm MBO})}, \qquad
L_{\rm MBO} = r_{\rm MBO}\sqrt{\frac{1-f(r_{\rm MBO})}{f(r_{\rm MBO})}}.
\label{eq:mbo-eq}
\end{equation}
The ISCO requires the inflection condition $\partial_r^2 V_{\rm eff}=0$, which combines with the circular orbit condition to give
\begin{equation}
r\,f(r)\,f''(r) - 2r\left[f'(r)\right]^{2} + 3\,f(r)f'(r) = 0.
\label{eq:isco-eq}
\end{equation}
The root $r_{\rm ISCO}$ determines the characteristic angular momentum and energy,
\begin{equation}
L_{\rm ISCO} = \sqrt{\frac{r_{\rm ISCO}^{3}\,f'(r_{\rm ISCO})}{2f(r_{\rm ISCO})-r_{\rm ISCO}f'(r_{\rm ISCO})}}, \qquad
E_{\rm ISCO} = \sqrt{\frac{2\,f(r_{\rm ISCO})^{2}}{2f(r_{\rm ISCO})-r_{\rm ISCO}f'(r_{\rm ISCO})}}.
\label{eq:LISCO_EISCO}
\end{equation}
For the lapse function in Eq.~\eqref{eq:lapse}, $f'(r) = 2MrN(r)/D(r)^{2}$ with $N(r) \equiv r^{3} - \alpha r - 2\alpha\gamma M$, which correctly reduces to $2M/r^{2}$ as $\alpha\to0$. In the Schwarzschild limit, Eqs.~\eqref{eq:mbo-eq}--\eqref{eq:LISCO_EISCO} recover the familiar values $r_{\rm MBO}=4M$, $r_{\rm ISCO}=6M$, $L_{\rm ISCO}=2\sqrt{3}M$, and $E_{\rm ISCO}=2\sqrt{2}/3$.

\begin{figure}[htbp]
\centering
\begin{subfigure}[b]{0.48\linewidth}
\centering
\includegraphics[width=\linewidth]{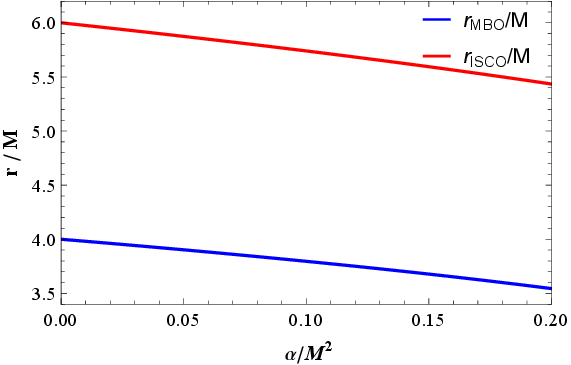}
\caption{Characteristic radii $r_{\rm MBO}/M$ (blue) and $r_{\rm ISCO}/M$ (red).}
\label{fig:r-vs-alpha}
\end{subfigure}
\hfill
\begin{subfigure}[b]{0.48\linewidth}
\centering
\includegraphics[width=\linewidth]{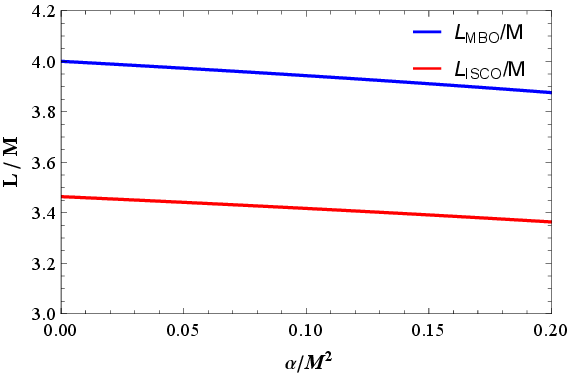}
\caption{Angular momenta $L_{\rm MBO}/M$ (blue) and $L_{\rm ISCO}/M$ (red).}
\label{fig:L-vs-alpha}
\end{subfigure}
\caption{Characteristic radii and angular momenta of the MBO and ISCO as functions of $\alpha/M^{2}$, for $\gamma=9/2$. Both quantities decrease monotonically from their Schwarzschild values towards the critical point $\alpha/M_c^2\simeq0.204$.}
\label{fig:MBO-ISCO-vs-alpha}
\end{figure}

Figures~\ref{fig:r-vs-alpha} and \ref{fig:L-vs-alpha} track the characteristic radii and angular momenta for $0\le\alpha/M^2\le0.20$. Both $r_{\rm MBO}/M$ and $r_{\rm ISCO}/M$ decrease monotonically from their Schwarzschild values as $\alpha/M^{2}$ grows, and the angular momenta follow the same trend. This indicates that the RG-improvement favors more tightly bound configurations, pushing the MBO and ISCO closer to the horizon at smaller angular momenta.

\begin{figure}[htbp]
\centering
\includegraphics[width=0.75\linewidth]{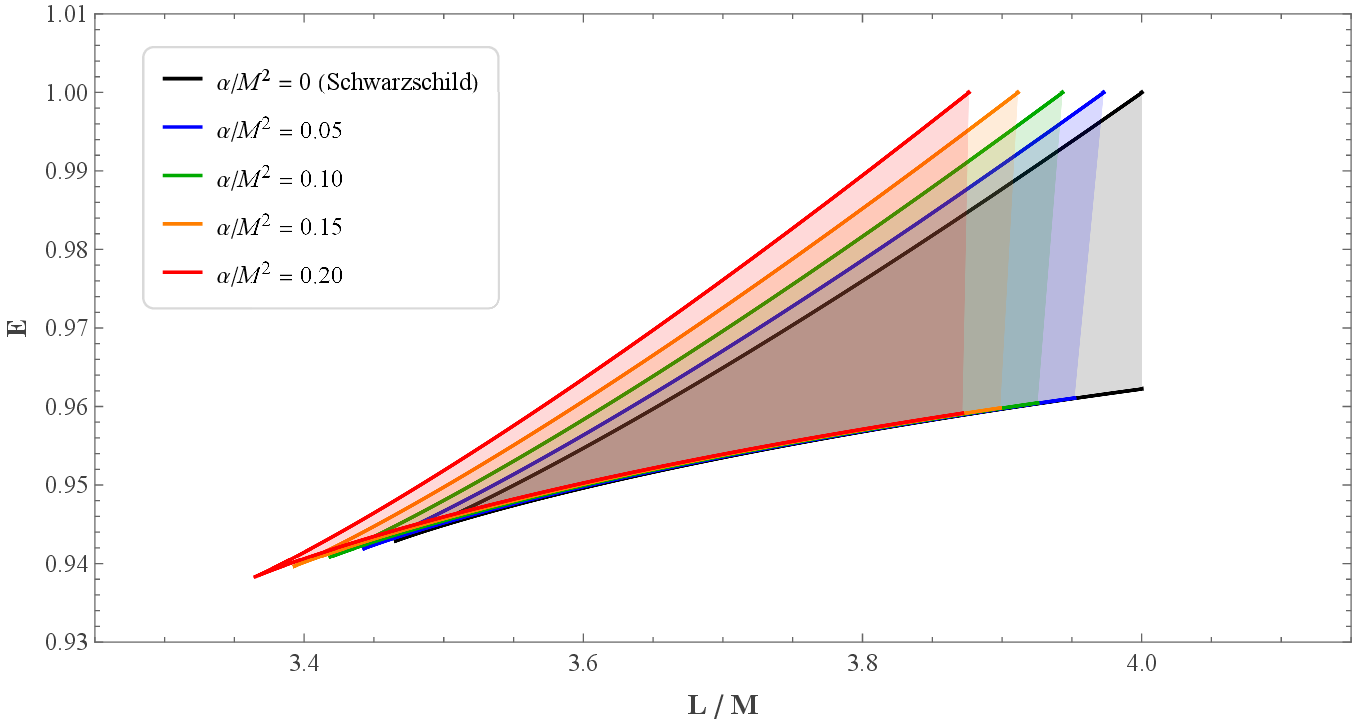}
\caption{Allowed parameter space for bound timelike motion in the $(E, L/M)$-plane for several values of $\alpha/M^{2}$, with $\gamma=9/2$. The boundary of each region is the locus of unstable circular orbits, and the allowed bound orbits lie below each curve.}
\label{fig:ELspace}
\end{figure}

Bound periodic orbits must satisfy $L \ge L_{\rm ISCO}$ and $E_{\rm ISCO} \le E \le E_{\rm MBO}=1$. Figure~\ref{fig:ELspace} maps this allowed parameter space in the $(E, L/M)$-plane. As $\alpha/M^{2}$ increases, the allowed region shifts systematically toward lower $L$ and $E$, consistent with the inward migration of the MBO and ISCO. This contraction defines the parameter window for the periodic orbits analyzed in Sec.~\ref{sec:periodic}.

\section{Periodic Orbits around the Bonanno--Reuter Black Hole}
\label{sec:periodic}

Periodic orbits provide a robust framework for probing the strong-field dynamics of black holes and for modeling gravitational-wave emission from extreme mass-ratio inspirals (EMRIs). In a static, spherically symmetric spacetime, a test particle on a bound geodesic is characterized by two fundamental frequencies: the radial frequency $\omega_r$ and the azimuthal frequency $\omega_\phi$. A trajectory is classified as periodic if the ratio of these frequencies is a rational number, allowing the orbit to close upon itself after a finite number of oscillations. Following the taxonomy introduced by Levin and Perez-Giz, each periodic orbit is uniquely labeled by a triplet of positive integers $(z, w, v)$, where $z$ is the zoom number (denoting the number of radial oscillations required for the orbit to close), $w$ is the whirl number (representing the number of full revolutions around the black hole per radial cycle), and $v$ is the vertex number (specifying the ordering of successive leaves). The frequency ratio is parameterized by the rational number $q$ as
\begin{equation}
q \equiv \frac{\omega_\phi}{\omega_r} - 1 = w + \frac{v}{z}.
\end{equation}
For a given value of the asymptotically safe gravity (ASG) parameter $\alpha/M^2$, periodic orbits are strictly confined to the bound region defined in Sec.~\ref{sec:mbo-isco}, satisfying $E_{\rm ISCO} \le E \le 1$ and $L \ge L_{\rm ISCO}$. Using the radial and azimuthal equations of motion, $q$ can be expressed explicitly as
\begin{equation}
q = \frac{1}{\pi} \int_{r_1}^{r_2} \frac{L}{r^2 \sqrt{E^2 - V_{\rm eff}(r)}} \, dr - 1,
\end{equation}
where $r_1$ and $r_2$ are the turning points (periastron and apastron) determined by the condition $E^2 = V_{\rm eff}(r)$. For specified pairs of specific energy $E$ and angular momentum $L$, this integral is evaluated numerically using standard quadrature methods.

To systematically explore the bound parameter space, it is convenient to fix one orbital constant to a representative average value. We define the average angular momentum $L_{\rm av} = (L_{\rm MBO} + L_{\rm ISCO})/2$ and the average energy $E_{\rm av} = (1 + E_{\rm ISCO})/2$. By solving the condition $q(E) = w + v/z$ with $L = L_{\rm av}$, we obtain the required periodic energies for different topological configurations $(z, w, v)$, which are tabulated in Table~\ref{tab:periodic_energies_ordered}. Similarly, solving $q(L) = w + v/z$ with $E = E_{\rm av}$ yields the corresponding angular momenta, presented in Table~\ref{tab:periodic_L_no_w3}. Across all configurations, increasing the ASG parameter $\alpha/M^2$ systematically reduces the energy required to sustain a given periodic orbit at fixed $L_{\rm av}$, and correspondingly lowers the required angular momentum at fixed $E_{\rm av}$. This indicates that the running Newton coupling modifies the effective potential such that periodic configurations are supported at lower specific energies and angular momenta.

\begin{table}[htbp]
    \centering
    \setlength{\arrayrulewidth}{0.5pt}
    \renewcommand{\arraystretch}{1.2}
    \scriptsize
    \setlength{\tabcolsep}{3pt}
    \resizebox{\textwidth}{!}{%
    \begin{tabular}{cccccccccc}
    \toprule
    $\alpha/M^2$ & $L_{\text{av}}$ & $E(1,1,0)$ & $E(1,2,0)$ & $E(2,1,1)$ & $E(2,2,1)$ & $E(3,1,2)$ & $E(3,2,2)$ & $E(4,1,3)$ & $E(4,2,3)$ \\
    \midrule
    0.00 & 3.732051 & 0.965425 & 0.968383 & 0.968027 & 0.968434 & 0.968225 & 0.968438 & 0.968285 & 0.968440 \\
    0.05 & 3.706987 & 0.964608 & 0.967856 & 0.967452 & 0.967917 & 0.967675 & 0.967922 & 0.967743 & 0.967923 \\
    0.10 & 3.680216 & 0.963669 & 0.967273 & 0.966807 & 0.967347 & 0.967063 & 0.967353 & 0.967142 & 0.967355 \\
    0.15 & 3.651365 & 0.962567 & 0.966620 & 0.966072 & 0.966711 & 0.966369 & 0.966719 & 0.966462 & 0.966721 \\
    0.20 & 3.619895 & 0.961236 & 0.965871 & 0.965209 & 0.965988 & 0.965563 & 0.966000 & 0.965676 & 0.966003 \\
    \bottomrule
    \end{tabular}%
    }%
    \caption{Periodic-orbit energies $E(z,w,v)$ for the Bonanno--Reuter black hole as functions of $\alpha/M^{2}$ with fixed average angular momentum $L=L_{\rm av}$, obtained by solving $q(E)=w+v/z$.}
    \label{tab:periodic_energies_ordered}
\end{table}

\begin{table}[htbp]
    \centering
    \setlength{\arrayrulewidth}{0.5pt}
    \renewcommand{\arraystretch}{1.2}
    \scriptsize
    \setlength{\tabcolsep}{3pt}
    \resizebox{\textwidth}{!}{%
    \begin{tabular}{cccccccccc}
    \toprule
    $\alpha/M^2$ & $E_{\rm av}$ & $L(1,1,0)$ & $L(1,2,0)$ & $L(2,1,1)$ & $L(2,2,1)$ & $L(3,1,2)$ & $L(3,2,2)$ & $L(4,1,3)$ & $L(4,2,3)$ \\
    \midrule
    0.00 & 0.971405 & 3.784077 & 3.759377 & 3.762349 & 3.758960 & 3.760687 & 3.758927 & 3.760186 & 3.758917 \\
    0.05 & 0.970920 & 3.760464 & 3.733984 & 3.737282 & 3.733501 & 3.735452 & 3.733462 & 3.734895 & 3.733450 \\
    0.10 & 0.970387 & 3.735509 & 3.706880 & 3.710592 & 3.706311 & 3.708552 & 3.706263 & 3.707923 & 3.706248 \\
    0.15 & 0.969796 & 3.708990 & 3.677709 & 3.681958 & 3.677020 & 3.679649 & 3.676959 & 3.678927 & 3.676939 \\
    0.20 & 0.969132 & 3.680614 & 3.645960 & 3.650938 & 3.645095 & 3.648271 & 3.645014 & 3.647422 & 3.644988 \\
    \bottomrule
    \end{tabular}%
    }%
    \caption{Periodic-orbit angular momenta $L(z,w,v)$ for the Bonanno--Reuter black hole as functions of $\alpha/M^{2}$ with fixed average energy $E=E_{\rm av}=(1+E_{\rm ISCO})/2$, obtained by solving $q(L)=w+v/z$.}
    \label{tab:periodic_L_no_w3}
\end{table}

\begin{figure}[htbp]
\centering
\begin{subfigure}[b]{0.48\linewidth}
    \centering
    \includegraphics[width=\linewidth]{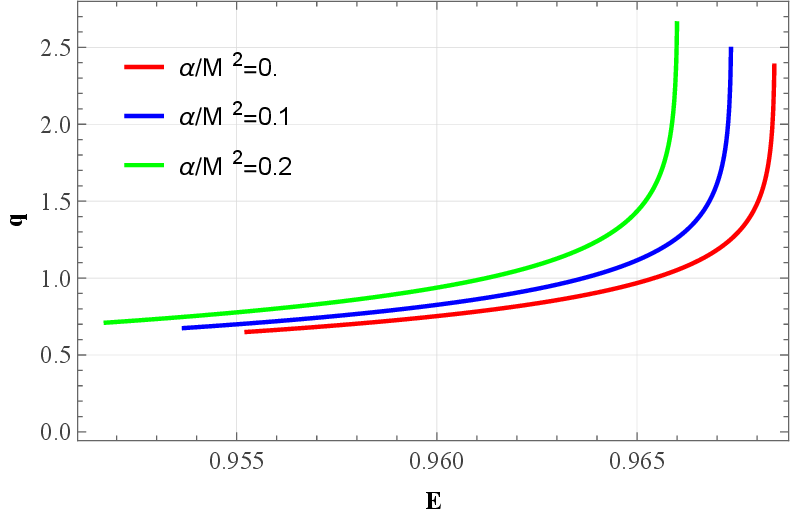}
    \caption{$q$ vs. $E$}
    \label{fig:q-vs-E}
\end{subfigure}
\hfill
\begin{subfigure}[b]{0.48\linewidth}
    \centering
    \includegraphics[width=\linewidth]{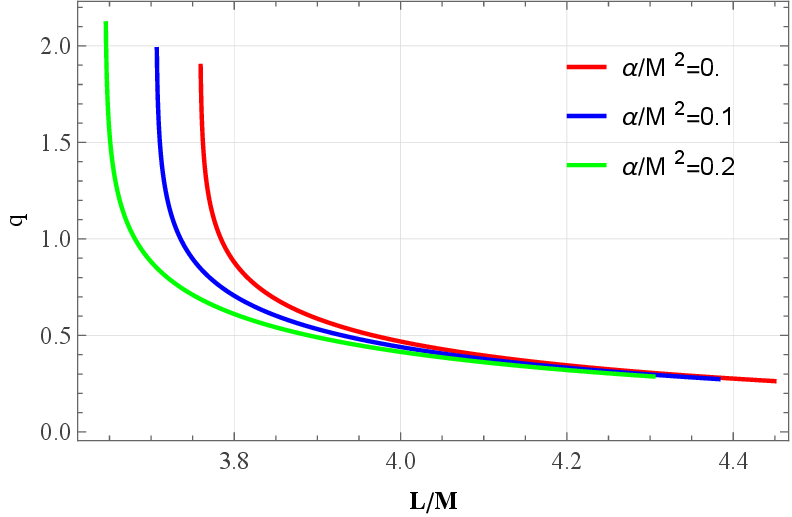}
    \caption{$q$ vs. $L/M$}
    \label{fig:q-vs-L}
\end{subfigure}
\caption{The rational number $q$ as a function of the orbital parameters for periodic bound orbits in the Bonanno--Reuter spacetime, with $\gamma=9/2$ fixed. \emph{Left:} $q$ as a function of the orbital energy $E$, with angular momentum fixed at $L=L_{\rm av}\equiv(L_{\rm ISCO}+L_{\rm MBO})/2$ for each value of $\alpha/M^{2}$. \emph{Right:} $q$ as a function of the angular momentum $L$, with energy fixed at $E=E_{\rm av}\equiv(1+E_{\rm ISCO})/2$. Increasing $\alpha/M^{2}$ systematically displaces both curves toward lower energies and angular momenta, reflecting the inward migration of the MBO and ISCO radii.}
\label{fig:q-vs-E-and-L}
\end{figure}

Figure~\ref{fig:q-vs-E-and-L} illustrates the dependence of the rational number $q$ on the orbital energy $E$ and angular momentum $L$. Crucially, increasing $\alpha/M^2$ shifts both the $q(E)$ and $q(L)$ curves systematically toward lower energies and angular momenta. This behavior is a direct kinematic manifestation of the inward migration of the MBO and ISCO radii discussed in Sec.~\ref{sec:mbo-isco}: as the quantum-gravity correction strengthens, the accessible bound-orbit window contracts, requiring a correspondingly lower $E$ or $L$ to achieve a specific rational $q$.

\begin{figure}[htbp]
\centering
\includegraphics[width=0.85\linewidth]{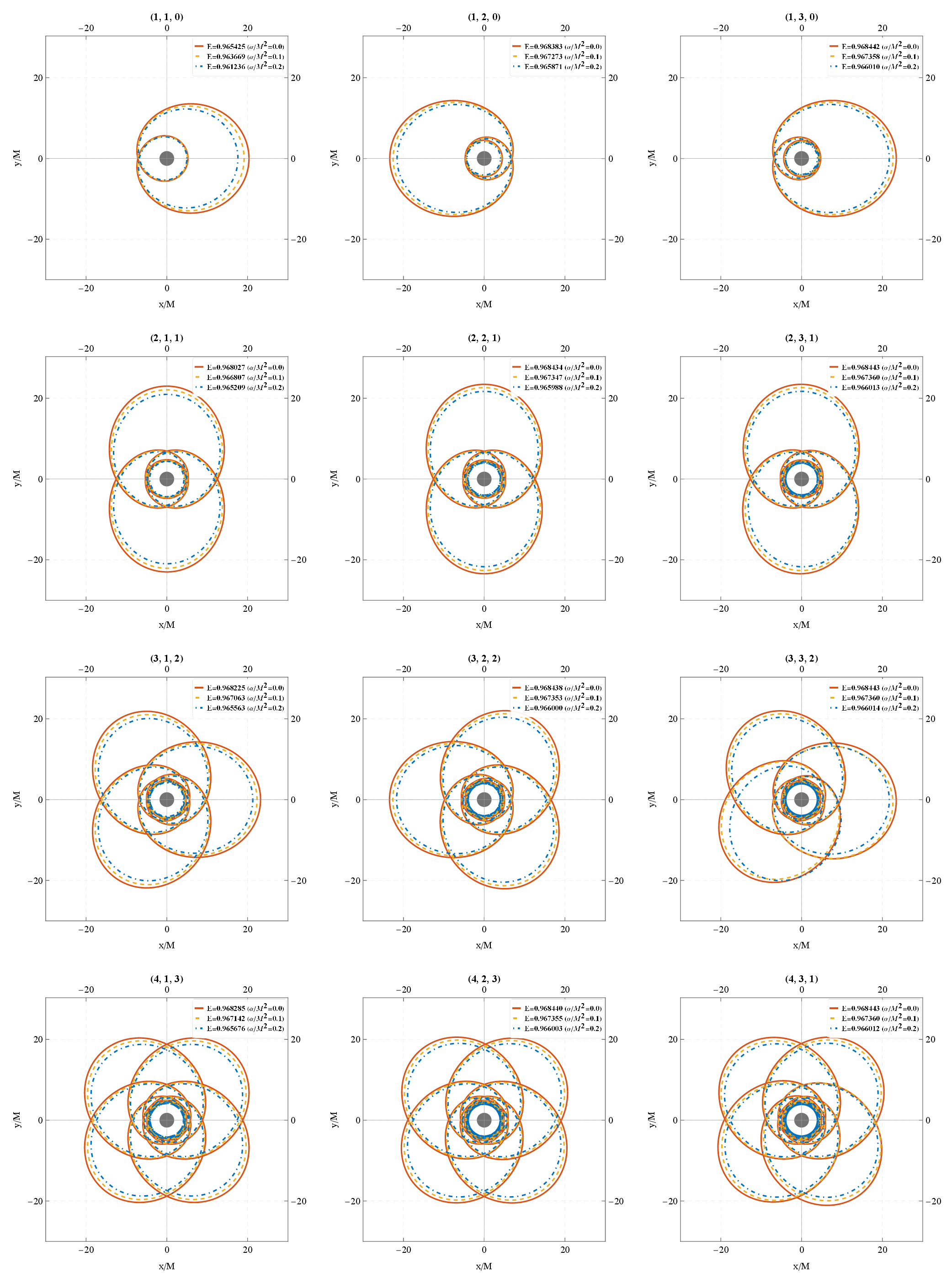}
\caption{Periodic timelike orbits around a Bonanno--Reuter regular black hole for various $(z, w, v)$ modes with $M = 1$ and $\gamma = 9/2$. In all panels, the angular momentum is fixed at $L_{\text{av}} = (L_{\text{MBO}} + L_{\text{ISCO}})/2$. Solid, dashed, and dot-dashed curves represent quantum corrections $\alpha/M^2 = 0.0$, $0.1$, and $0.2$, respectively, with their corresponding periodic energies $E$ listed in the legends. The central gray disk indicates the event horizon.}
\label{fig:periodic-orbits-E}
\end{figure}

\begin{figure}[htbp]
\centering
\includegraphics[width=0.85\linewidth]{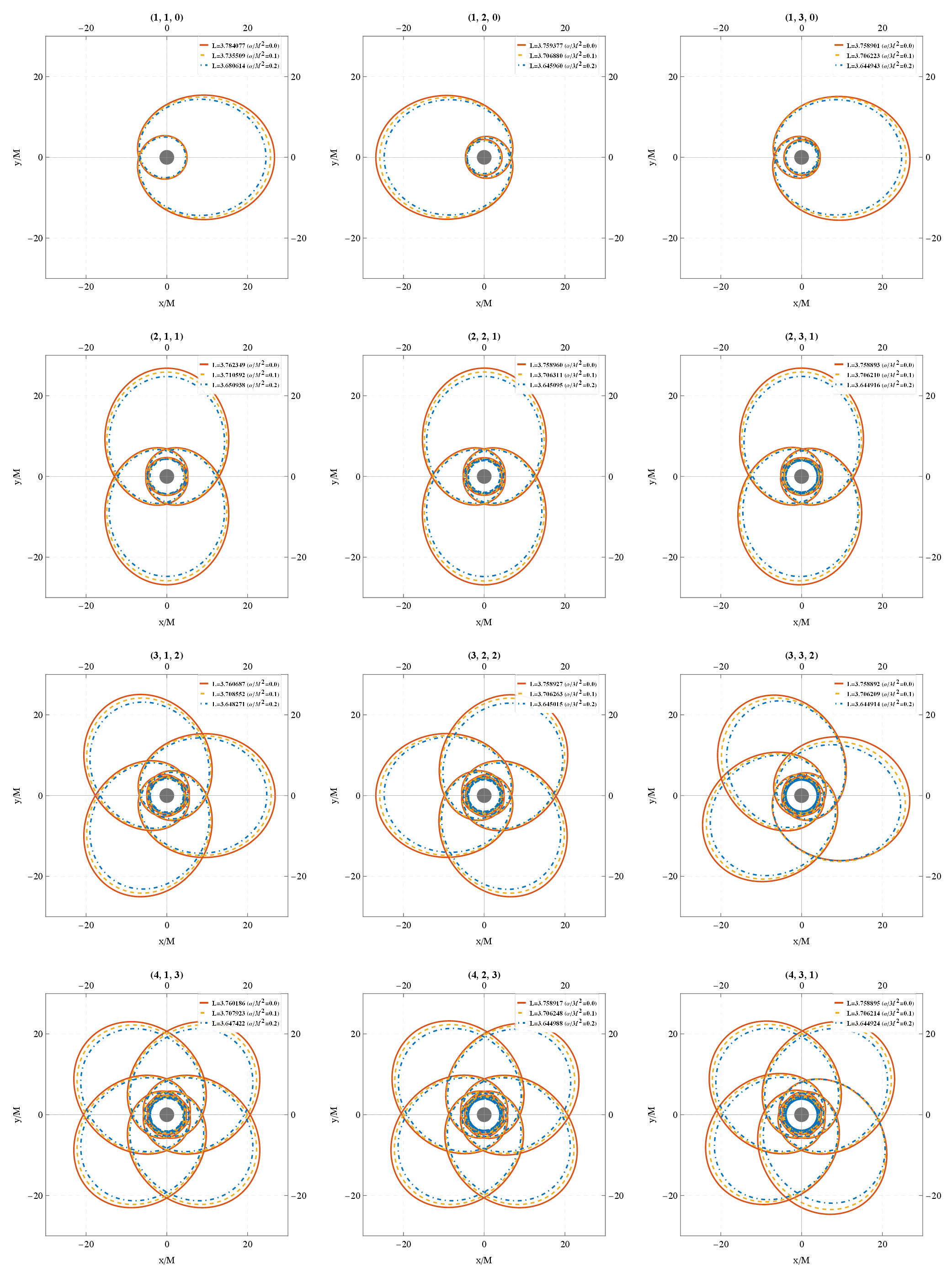}
\caption{Periodic timelike orbits around a Bonanno--Reuter regular black hole for various $(z, w, v)$ modes with $M = 1$ and $\gamma = 9/2$. In all panels, the energy is fixed at $E_{\text{av}} = (E_{\text{MBO}} + E_{\text{ISCO}})/2$. Solid, dashed, and dot-dashed curves represent quantum corrections $\alpha/M^2 = 0.0$, $0.1$, and $0.2$, respectively, with their corresponding periodic angular momenta $L$ listed in the legends. The central gray disk indicates the event horizon.}
\label{fig:periodic-orbits-L}
\end{figure}

The spatial trajectory of a periodic orbit in the equatorial $(x,y)$ plane is reconstructed by numerically integrating the radial and azimuthal equations of motion over one complete radial period. The resulting closed curves display the characteristic zoom-whirl morphology: the particle spends the majority of its time in the zoom phase, executing slow large-scale radial excursions near the apastron, punctuated by the whirl phase, during which it completes several rapid, nearly circular revolutions close to the black hole. This behavior is prominently featured in the representative trajectories displayed in Figs.~\ref{fig:periodic-orbits-E} and \ref{fig:periodic-orbits-L} for fixed $L_{\rm av}$ and fixed $E_{\rm av}$, respectively. Increasing the zoom number $z$ at fixed $w$ produces progressively more intricate multi-lobed patterns, whereas increasing the whirl number $w$ at fixed $z$ increases the density of near-circular revolutions executed near the periastron before the particle zooms back out. This topological structure is robust and model-independent, governed entirely by the rational value of $q$.

The specific imprint of the ASG parameter $\alpha/M^2$ on the orbital geometry is a systematic, albeit mild, inward contraction. As $\alpha/M^2$ increases from $0$ to $0.2$, the apastron excursions shrink appreciably, and the inner whirling portions of the trajectories draw closer to the event horizon. This contraction is the direct orbital manifestation of the shift in the characteristic radii $r_{\rm MBO}$ and $r_{\rm ISCO}$. Consequently, the trajectories for different $\alpha/M^2$ remain nearly superimposed for low-$w$ configurations, becoming progressively more distinguishable only for the more tightly wound, high-$w$ orbits.

In summary, the periodic orbits around the Bonanno--Reuter black hole adhere to the standard zoom-whirl classification and $q(E,L)$ phenomenology. The asymptotically safe correction leaves a distinct signature through a systematic inward contraction of the orbits and a corresponding downward shift of the periodic energies and angular momenta, providing the necessary geometric foundation for the gravitational waveform analysis presented in the subsequent section.

\section{Gravitational Waveform from Periodic Orbits}
\label{sec:waveforms}

We investigate the gravitational radiation emitted by a stellar-mass compact object moving along the periodic trajectories established in Sec.~\ref{sec:periodic}. The system is modeled as an extreme mass-ratio inspiral (EMRI), comprising a compact object of mass $m$ orbiting a supermassive Bonanno--Reuter primary black hole of mass $M$. The waveform calculation is performed within the numerical kludge framework, which combines the numerically integrated geodesic trajectory with the standard quadrupole formula. In the adiabatic limit, the radiation-reaction timescale is assumed to be much longer than the orbital period; thus, the orbital energy and angular momentum are treated as conserved over a single radial cycle, and the backreaction on the orbital motion is neglected.

Within this framework, the metric perturbation in the wave zone is derived from the symmetric trace-free (STF) mass quadrupole tensor, given by
\begin{equation}
h_{ij} = \frac{4\mu M}{D_L} \left( v_i v_j - \frac{m}{M} n_i n_j \right),
\end{equation}
where $\mu = Mm/(M+m)^2$ is the symmetric mass ratio, $D_L$ is the luminosity distance to the source, and $n_i$ and $v_i$ represent the radial unit vector and the velocity components of the orbiting body, respectively. To obtain the observable waveform, this metric perturbation is projected onto a transverse-traceless frame adapted to the detector. Introducing the orthonormal basis vectors
\begin{equation}
\begin{aligned}
\mathbf{e}_X &= (\cos \zeta, -\sin \zeta, 0), \\
\mathbf{e}_Y &= (\cos \iota \sin \zeta, \cos \iota \cos \zeta, -\sin \iota), \\
\mathbf{e}_Z &= (\sin \iota \sin \zeta, \sin \iota \cos \zeta, \cos \iota),
\end{aligned}
\end{equation}
where $\iota$ denotes the orbital inclination and $\zeta$ specifies the orientation of the orbit relative to the observer, the two independent polarization modes are expressed as
\begin{equation}\label{eq:hplus}
h_+ = -\frac{2\mu M^2}{D_L r} \left(1 + \cos^2 \iota\right) \cos(2\phi + 2\zeta),
\end{equation}
\begin{equation}\label{eq:hcross}
h_\times = -\frac{4\mu M^2}{D_L r} \cos \iota \sin(2\phi + 2\zeta),
\end{equation}
with $r$ and $\phi$ denoting the instantaneous orbital radius and phase. These quantities are obtained directly from the numerical integration of the timelike geodesic equations governed by the Bonanno--Reuter metric. For our numerical simulations, we adopt geometrized units and fix the mass scales at $M \sim 10^7 M_\odot$ and $m \sim 10 M_\odot$, with the source placed at a luminosity distance of $D_L = 200$ Mpc. The orientation angles are fixed at $\iota = \zeta = \pi/4$.

\begin{figure}[htbp]
\centering
\includegraphics[width=0.85\linewidth]{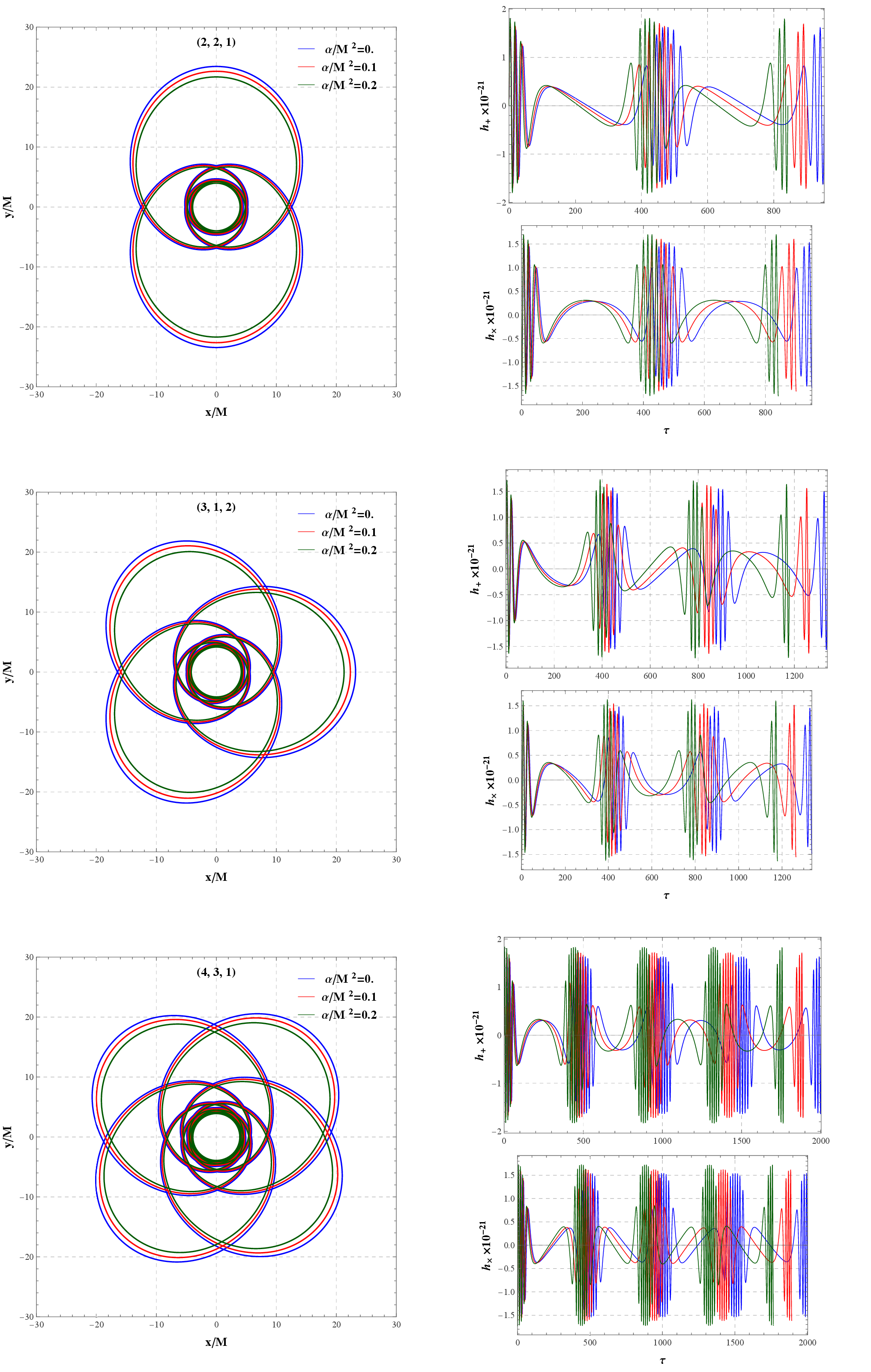}
\caption{Periodic orbits and the corresponding gravitational-wave polarizations around a Bonanno--Reuter black hole. From top to bottom, the rows represent the $(2,2,1)$, $(3,1,2)$, and $(4,3,1)$ orbital configurations. Left panels show the trajectories for $\alpha/M^{2}=0.0,\,0.1,\,0.2$ (solid, dashed, and dot-dashed curves), while right panels display the corresponding $h_{+}$ and $h_{\times}$ waveforms. The central gray disk indicates the event horizon.}
\label{fig:waveforms}
\end{figure}

Figure~\ref{fig:waveforms} displays the periodic trajectories in the equatorial $(x,y)$ plane alongside the corresponding time-domain gravitational-wave polarizations, $h_+$ and $h_\times$, for three representative orbital configurations: $(2,2,1)$, $(3,1,2)$, and $(4,3,1)$. The waveforms faithfully encode the characteristic zoom-whirl morphology of the underlying geodesics. The slowly varying, low-amplitude segments of the signal correspond to the zoom phase, during which the compact object traverses the outer regions of the orbit. Conversely, transient high-frequency bursts are radiated during the whirl phase, when the object executes multiple near-circular revolutions in the strong-field region close to the black hole. This burst-like behavior becomes increasingly prolonged for configurations with higher whirl numbers, as is evident in the $(4,3,1)$ orbit.

The specific imprint of the asymptotically safe gravity (ASG) parameter $\alpha/M^2$ on the gravitational-wave signal is both systematic and topology-dependent. As $\alpha/M^2$ increases, the periodic orbits contract inward toward the event horizon (Sec.~\ref{sec:periodic}). This geometric contraction directly translates into a leftward phase shift in the time-domain waveforms, reflecting a reduction in the orbital period. Because both polarizations scale as $1/r$ [see Eqs.~(\ref{eq:hplus}) and (\ref{eq:hcross})], the smaller periastron distance reached at fixed topology $(z,w,v)$ directly translates into a mild \emph{enhancement} of the peak amplitudes during periastron passage as $\alpha/M^2$ increases, despite the concurrent decrease in the periodic energy and angular momentum required to sustain that orbit.

Crucially, the sensitivity of the waveform to the ASG correction is dictated by the orbital topology. For the simpler $(2,2,1)$ configuration, the waveforms corresponding to $\alpha/M^2 = 0.0, 0.1, 0.2$ remain nearly superimposed, indicating a relatively weak dependence on the quantum-gravity correction. In contrast, the more complex, tightly wound $(4,3,1)$ orbit exhibits a clear separation between the waveforms for different values of $\alpha/M^2$. This enhanced sensitivity arises because orbits with higher whirl numbers spend a disproportionately longer time in the deep strong-field regime, precisely where the RG-improvement to the metric is most significant. Consequently, short-distance quantum-gravity modifications intrinsic to the Bonanno--Reuter geometry produce distinct, topology-dependent signatures in EMRI gravitational-wave signals, acting in the opposite direction to environmental effects that tend to inflate the orbital scales.

\section{Summary and Conclusions}

We analyzed the timelike geodesics, periodic orbits, and gravitational-wave signals for extreme mass-ratio inspirals (EMRIs) in the spacetime of a Bonanno--Reuter regular black hole, a geometry derived from Asymptotically Safe Gravity (ASG) where a running Newtonian coupling replaces the Schwarzschild singularity with a regular de Sitter core. Strong-field dynamics is governed entirely by the dimensionless parameter $\alpha/M^2$, which is bounded above by the extremal threshold $\alpha/M_c^2 \simeq 0.204$.

Analysis of the effective potential shows that increasing $\alpha/M^2$ monotonically decreases the radii and angular momenta of both the marginally bound orbit (MBO) and the innermost stable circular orbit (ISCO). Consequently, the region of bound motion in the $(E,L)$ plane shifts toward lower energies and angular momenta, indicating that the ASG correction promotes more tightly bound configurations.

The rational frequency ratio $q = w + v/z$ classifies bound trajectories, revealing that periodic orbits follow the standard zoom-whirl taxonomy. Systematically increasing $\alpha/M^2$ shifts the $q(E)$ and $q(L)$ profiles to lower values, producing a mild inward contraction of the orbital trajectories that is most pronounced for tightly wound, high-whirl configurations.

Using the numerical kludge framework, we computed the plus and cross gravitational-wave polarizations generated by a stellar-mass compact object orbiting a supermassive primary. The waveforms accurately capture the zoom-whirl morphology intrinsic to the system. As $\alpha/M^2$ increases, the orbital period decreases, resulting in a phase shift toward earlier times in the time domain. Concurrently, inward orbital contraction reduces the periastron distance. Since waveform amplitudes scale as $1/r$, this geometric effect produces a mild enhancement of peak amplitudes at periastron, even though sustaining the same topology requires lower orbital energies. The waveform sensitivity to the ASG correction depends on the orbital topology, as high-whirl orbits persist longer in the strong-field region and exhibit significantly more pronounced deviations.

Intrinsic quantum-gravity modifications to the compact object produce distinct, topology-dependent EMRI signatures, as these results demonstrate. The ASG correction contracts orbits and lowers their characteristic energies, acting in the opposite direction to environmental effects that inflate orbital scales. Future space-based detectors like LISA, Taiji, and TianQin could effectively probe these short-distance modifications, especially by observing high-whirl EMRI systems.

\end{document}